\documentclass{article}
\usepackage{amsmath,amssymb}
\usepackage[dvips]{graphicx}

\newtheorem{thm}{Theorem}[section]
\newtheorem{lem}[thm]{Lemma}
\newtheorem{cor}[thm]{Corollary}
\newtheorem{pro}[thm]{Proposition}
\newtheorem{rem}[thm]{Remark}

\newcommand{\ZM}{\mathbb{Z}}
\newcommand{\ZMP}{\mathbb{Z}_{+}}

\newcommand{\CM}{\mathbb{C}}

\newcommand{\fp}{\widetilde{f}^{(+)}}
\newcommand{\fm}{\widetilde{f}^{(-)}}
\newcommand{\lp}{\widetilde{\lambda}^{(+)}}
\newcommand{\lm}{\widetilde{\lambda}^{(-)}}
\newcommand{\ket}[1]{|#1\rangle}

\title{Limit measures of inhomogeneous discrete-time \\ quantum walks in one dimension
}
\author
{
{Norio KONNO,$^{1, }$\footnote{E-mail: konno@ynu.ac.jp}
 \quad Tomasz {\L}UCZAK,$^{2, }$\footnote{E-mail: tomasz@amu.edu.pl} 
 \quad Etsuo SEGAWA$^{3, }$\footnote{E-mail: segawa@stat.t.u-tokyo.ac.jp}}\\
{\scriptsize {}$^{1}$ \textit{Department of Applied Mathematics, Faculty of Mathematics and Computer Science,}}\\  
{\scriptsize \textit{Yokohama National University, Hodogaya, Yokohama, 240-8501, Japan}}\\
{}
{\scriptsize {}$^{2}$ \textit{Department of Discrete Mathematics, Faculty of Engineering,}}\\  
{\scriptsize \textit{Adam Mickiewicz University, Umultowska 87, 61-614, Pozna\'n, Poland}}\\
{}
{\scriptsize {}$^{3}$ \textit{Department of Mathematical Informatics, }}\\ 
{\scriptsize \textit{University of Tokyo, Bunkyo, Tokyo, 113-8656, Japan}}\\
}
\date{}
\begin{document}
\maketitle
\noindent
\begin{small}
\textbf{Abstract. }
We treat three types of measures of the quantum walk (QW) with the spatial perturbation at the origin, 
which was introduced by \cite{KonnoLocal}: 
time averaged limit measure, weak limit measure, and stationary measure. 
From the first two measures, we see a coexistence of the ballistic and localized behaviors in the walk 
as a sequential result following \cite{KonnoLocal,KonnoRE}. 
We propose a universality class of QWs with respect to weak limit measure. 
It is shown that typical spatial homogeneous QWs with ballistic spreading belong to the universality class. 
We find that the walk treated here with one defect also belongs to the class. 
We mainly consider the walk starting from the origin. 
However when we remove this restriction, we obtain a stationary measure of the walk. 
As a consequence, by choosing parameters in the stationary measure, 
we get the uniform measure as a stationary measure of the Hadamard walk and 
a time averaged limit measure of the walk with one defect respectively. 
\end{small}
\section{Introduction}
\quad This paper is a sequential work following \cite{KonnoRE} (2009) and \cite{KonnoLocal} (2010) by the first author. 
Localization is considered as a typical property of discrete-time QWs 
beginning from \cite{TregennaEtAl,IKK} in the numerical simulations. 
Let $X_n$ be the walk at time $n$. In this paper, we say that 
the walk exhibits localization at position $x\in \mathbb{Z}$ 
if and only if its time averaged limit measure is exactly positive, {i.e.,} 
\begin{equation}
\lim_{T\to \infty} \frac{1}{T} \sum_{n=0}^{T-1} P(X_n=x)>0. 
\end{equation} 
The other definitions of localization have been discussed in \cite{ShikanoKatsura,JoyeP,AhlbrechtP}. 
There are three classes of QWs to see localization of one-dimensional QW. \\
\quad The first class is QWs with the spatial perturbation. 
Each quantum coin fixed through the steps depends on each position. 
By numerical simulations, localization occurs with a quasi-periodic perturbation in \cite{W}. 
A QW with one defect was introduced by \cite{KonnoLocal}. 
It was shown that the QW tends to exhibit a localization in even small perturbation. 
Cantero \textit{et al.} \cite{CGMV2} presented an explicit expression of necessary and sufficient condition 
for localization in QWs with one defect and found an importance of the initial coin state for the localization, 
from the view point of the spectral theory based on, for example, \cite{BS,CGMV,KS2010}. 
Some periodic and quasi-aperiodic perturbations implies 
bounded or unbounded QWs \cite{ShikanoKatsura,LindenSharam}.
Recently, it was shown in \cite{JoyeP,AhlbrechtP} that 
QW in a (spatial) random environment given by a general distribution exhibits localization. 
However, remark that there is also a QW with spatial random environment 
which has just a ballistic spreading without localization \cite{KonnoRE}. \\
\quad The second class is QWs with the temporal perturbation. 
The quantum coin at each time step is updated. 
By using a combinatorial analysis introduced by \cite{Konno2002,Konno2005}, 
Konno\cite{KonnoDis} (2005) showed the distribution averaged by the environment for 
temporal disordered QWs including the models numerically simulated by \cite{RibeiroEtAl,MackayEtAl}. 
The temporal perturbation in the above QWs implies a central limit theorem. 
It is reported in \cite{JoyeT,AhlbrechtT} that a more general setting for a temporal perturbation also gives 
a similar statement to the Konno's result in \cite{KonnoDis}. 
It seems that a temporal disorder tends to induce the diffusive spreading from the above models \cite{KonnoDis,JoyeT,AhlbrechtT}. 
However, recently 
Chandrashekar \cite{Chandrashekar} showed that his class of temporal perturbation models exhibits localization 
through numerical simulations. 
A temporal one defect walk studied in \cite{Machida} exhibits also localization. \\
\quad The last class is QWs with multi-coin state introduced by \cite{IK}. 
Increasing the coin state also exhibits localization. 
As is first suggested by Inui \textit{et al.,} \cite{IKS} (2005), 
we need two kinds of limit theorems in this walk to describe the limit behaviors. 
The first one corresponds to localization. 
However, we can see that the summation of the asymptotic measure over the all position, $C$, is strictly smaller than one. 
The second one is the weak convergence theorem corresponding to the missing part $1-C$. 
Each of weak limit measures $\mu$ for the multi-state coin QW appeared in \cite{IKS,FKK,MKK,SK,KM,LP} has a common expression. 
Therefore, we present a universality class of QWs whose weak limit measures have the expression. 
Concretely, the weak limit measure is given in the following way: 
there exist a rational polynomial $w(x)$, and real numbers $C\in [0,1)$, $r\in (0,1)$ 
such that $\mu(dx)=C\delta_0(x)+w(x)f_K(x;r)dx$, where 
\begin{equation}\label{univ} f_K(x;r)= \frac{\sqrt{1-r^2}}{\pi (1-x^2)\sqrt{r^2-x^2}}I_{(-r,r)}(x), \end{equation}
with $I_A(x)=1$, $(x\in A)$, $=0$, $(x\notin A)$. 
Remark that \[ 1-C=\int_{-\infty}^{\infty}w(x)f_K(x;r).\]
\quad 
It was shown that the QW studied in \cite{KonnoRE} belongs to the universality class $(C=0)$. 
However, it is not known that the QW in \cite{KonnoLocal} belongs to the class. 
In this paper, we give an affirmative answer to this problem, indeed $C\in[0,1)$, 
by using a path counting method. 
Recently, Liu and Petulante \cite{ChaobinPetulante} treated limit behaviors determined by 
bistochastic quantum operations on a finite system. On the other hand, for an infinite system, 
we obtain stationary and time averaged limit measures which belong to categories (2) and (4) in their notation. \\
\quad This paper is organized as follows. 
Section 2 gives the definition of the QW treated here. Moreover we define three measures of the walk: 
stationary measure, time averaged limit measure, and weak limit measures. 
We present the generating function of weight of passage and the time averaged measure of the walk in Sect. 3. 
Section 4 is devoted to the weak limit measure. 
We show that this walk belongs to the universality class we introduced. 
In the final section, we give an expression for the stationary measure 
when we remove the restriction of the initial state starting from the origin. 
\section{Definition of the walk}
In this section, we give the definition of the space-inhomogeneous two-state QW on $\ZM$ considered here, where $\ZM$ is the set of integers. 
The discrete-time QW is a quantum version of the classical random walk with additional degree of freedom called chirality. 
The chirality takes values left and right, and it means the direction of the motion of the walker. 
At each time step, if the walker has the left chirality, it moves one step to the left, and if it has the right chirality, it moves one step to the right. 
Define
\begin{eqnarray*}
\ket{L} = 
\left[
\begin{array}{cc}
1 \\
0  
\end{array}
\right],
\qquad
\ket{R} = 
\left[
\begin{array}{cc}
0 \\
1  
\end{array}
\right],
\end{eqnarray*}
where $L$ and $R$ refer to the left and right chirality state, respectively.  

For the general setting, the time evolution of the walk is determined by a sequence of $2 \times 2$ unitary matrices, $\{ U_x : x \in \ZM \}$, where
\begin{align*}
U_x =
\left[
\begin{array}{cc}
a_x & b_x \\
c_x & d_x
\end{array}
\right],
\end{align*}
with $a_x,b_x,c_x,d_x \in \mathbb C$ and $\CM$ is the set of complex numbers.  
The subscript $x$ indicates the location. The matrices $U_x$ rotate the chirality before the displacement, which defines the dynamics of the walk. 
To describe the evolution of our model, we divide $U_x$ into two matrices:
\begin{eqnarray*}
P_x =
\left[
\begin{array}{cc}
a_x & b_x \\
0 & 0 
\end{array}
\right], 
\quad
Q_x=
\left[
\begin{array}{cc}
0 & 0 \\
c_x & d_x 
\end{array}
\right],
\end{eqnarray*}
with $U_x=P_x+Q_x$. The important point is that $P_x$ (resp. $Q_x$) represents that the walker moves to the left (resp. right) at position $x$ at each time step. 

For a given sequence $\{ (\omega_x, \widetilde{\omega}_x) : x \in \ZM \}$ with $\omega_x, \widetilde{\omega}_x \in [0, 2 \pi)$, let  
\begin{align}
U_x (\omega_x, \widetilde{\omega}_x) = 
\frac{1}{\sqrt{2}}
\left[
\begin{array}{cc}
e^{i \omega_x} & e^{i \widetilde{\omega}_x}  \\
e^{-i \widetilde{\omega}_x} & -e^{-i \omega_x} 
\end{array}
\right].
\label{seiko}
\end{align} 
Our previous papers \cite{KonnoLocal} and \cite{KonnoRE} treated the QWs determined by $U_x (\omega_x, 0)$ and $U_x (0, \widetilde{\omega}_x)$, respectively. 
Here we concentrate on a simple inhomogeneous model with one defect 
\[ U_x= U_0\;(x=0),\;\; U_x=U\;(x\neq 0). \]
In particular, our previous paper \cite{KonnoLocal} corresponds to the following special case: 
\begin{align}
U_0 = U_0 (0, \omega), \quad U_x = U_x (0,0) \quad \hbox{if} \> x \not= 0. 
\label{seikou}
\end{align}
So when $\omega \not=0$, our model is homogeneous except the origin. 
If $\omega = 0$, then this model becomes homogeneous and is equivalent to the {\it Hadamard walk} determined by the Hadamard gate $U_x = U_x (0,0) \equiv H$:
\begin{eqnarray*}
H=\frac{1}{\sqrt2}
\left[
\begin{array}{cc}
1 & 1 \\
1 &-1 
\end{array}
\right].
\end{eqnarray*}

Let $\Psi_n (x)$ denote the amplitude of space-inhomogeneous QW at position $x$ at time $n$ as follows: 
\begin{align*}
\Psi_n (x)
= \left[
\begin{array}{cc}
\Psi_n^{(L)} (x) \\
\Psi_n^{(R)} (x) 
\end{array}
\right].
\end{align*}
Then the definition of the evolution gives 
\begin{align*}
\Psi_{n+1} (x) = P_{x+1} \Psi_{n} (x+1) + Q_{x-1} \Psi_{n} (x-1), 
\end{align*}
that is,
\begin{align*}
\left[
\begin{array}{cc}
\Psi_{n+1}^{(L)} (x) \\
\Psi_{n+1}^{(R)} (x) 
\end{array}
\right]
=
\left[
\begin{array}{cc}
a_{x+1} \Psi_{n}^{(L)} (x+1) + b_{x+1} \Psi_{n}^{(R)} (x+1) \\
c_{x-1} \Psi_{n}^{(L)} (x-1) + d_{x-1} \Psi_{n}^{(R)} (x-1)
\end{array}
\right].
\end{align*}
Put 
\begin{align*}
\Psi_n = {}^T \left[ \ldots, \Psi_n^{(L)} (-1), \Psi_n^{(R)} (-1), \Psi_n^{(L)} (0), \Psi_n^{(R)} (0), \Psi_n^{(L)} (1), \Psi_n^{(R)} (1), \ldots \right]
\end{align*}
and
\begin{eqnarray*}
U^{(s)} =
\left[
\begin{array}{ccccccc}
\ddots & \vdots & \vdots & \vdots & \vdots & \vdots & \ldots \\
\ldots & O      & P_{-1} & O      & O   & O   & \ldots \\
\ldots & Q_{-2} & O     & P_{0} & O   & O   & \ldots \\
\ldots & O      & Q_{-1}   & O      & P_1 & O   & \ldots \\
\ldots & O      & O     & Q_0    & O   & P_2 & \ldots \\
\ldots & O      & O     & O      & Q_1 & O   & \ldots \\
\ldots & \vdots & \vdots & \vdots & \vdots & \vdots & \ddots
\end{array}
\right], 
\quad
\hbox{where}
\quad
O = 
\left[
\begin{array}{cc}
0 & 0 \\
0 & 0 
\end{array}
\right].
\end{eqnarray*}
Then the state of the QW at time $n$ is given by $\Psi_{n} = (U^{(s)})^n \Psi_{0}$ for any $n \ge 0$. 
The measure that our quantum walker exists in position $x$ at time $n$ with initial state $\Psi_{0}$ is defined by 
\begin{align*}
\mu_n (x) = \left|\left| \Psi_n (x) \right|\right|^2 = \left| \Psi^{(L)}_n (x) \right|^2 + \left| \Psi^{(R)}_n (x) \right|^2 \quad (x \in \ZM).
\end{align*}
In general, we define a map $\varphi: (\CM^2)^{\ZM} \to \ZMP^{\ZM}$ as follows: for $\Psi \in  (\CM^2)^{\ZM}$ with
\begin{align*}
\Psi = {}^T \left[ \ldots, 
\begin{bmatrix} \Psi^{(L)} (-1) \\ \Psi^{(R)} (-1) \end{bmatrix}, 
\begin{bmatrix} \Psi^{(L)} (0)  \\ \Psi^{(R)} (0) \end{bmatrix}, 
\begin{bmatrix} \Psi^{(L)} (1)  \\ \Psi^{(R)} (1) \end{bmatrix}, 
\ldots \right],
\end{align*}
let $\mu = \varphi (\Psi)$ with  
\begin{align*}
\mu (x) = \varphi \left( \Psi (x) \right) = \left| \Psi^{(L)} (x) \right|^2 + \left| \Psi^{(R)} (x) \right|^2 \quad (x \in \ZM).
\end{align*}
Note that $\mu_n (x) = \varphi \left( \Psi_n (x) \right)$. 
Define
\begin{align*}
\mathcal{M}_I
&= \left\{ \varphi \left( \Psi_0 \right) \in \ZMP^{\ZM}\setminus\{\boldsymbol{0}\} : 
\Psi_0 \in \CM^{\ZM} \> \hbox{satisfies} \> \varphi \left( (U^{(s)})^n \Psi_0 \right) = \varphi \left( \Psi_0 \right) \> \hbox{for any} \> n \ge 0. \right\}.
\end{align*} 
We call the element of $\mathcal{M}_I$ the stationary measure of the QW. 
We rewrite the element of $\mathcal{M}_I$ by $\mu_I^{(\Psi_0)}$ to emphasize ``stationary distribution" and 
the dependence on the initial state $\Psi_0$. 
Moreover, we introduce the time average of $\mu_n (x)$ and its limit:
\begin{align*}
\overline{\mu}_{T} (x) 
&= \frac{1}{T} \sum_{n=0}^{T-1} \mu_n (x), 
\\
\overline{\mu}_{\infty} (x) 
&= \lim_{T \to \infty} \overline{\mu}_{T} (x) = \lim_{T \to \infty} \frac{1}{T} \sum_{n=0}^{T-1} \mu_n (x).
\end{align*}
Put 
\begin{align*}
\overline{\mathcal{M}}_{\infty} 
&= \left\{ \overline{\mu}_{\infty} = \overline{\mu}_{\infty}^{(\Psi_0)} \in \ZMP^{\ZM}\setminus\{0\} : \Psi_0 \in \CM^{\ZM} \right\},
\end{align*}
where $\overline{\mu}_{\infty}^{(\Psi_0)}$ represents the dependence on the initial state $\Psi_0$. 
We call the element of $\overline{\mathcal{M}}_{\infty}$ the time-averaged limit measure of the QW. 
Obviously, we have $\mathcal{M}_I\subset \overline{\mathcal{M}}_\infty$. 
Put $\boldsymbol{\widetilde{\Phi}^{(o)}}=\{ \Psi_0\in (\mathbb{C}^2)^{\mathbb{Z}}: ||\Psi_0||^2=1 \}$. 
Let $X_n^{(\Psi_0)}$ be a random variable whose distribution is defined by $P(X_n^{(\Psi_0)}=x)=\mu_n(x)$ with the 
initial state $\Psi_0\in \boldsymbol{\widetilde{\Phi}^{(o)}}$. 
Let $\mu^{(Y)}$ denote the measure induced by a random variable $Y$.  
For $\Psi_0\in \boldsymbol{\widetilde{\Phi}^{(o)}}$, we put 
\begin{equation*}
\mathcal{M}_W(\Psi_0) = \left\{ \mu^{(Z^{(\Psi_0)})}: \frac{X_n^{(\Psi_0)}}{n}\Rightarrow Z^{(\Psi_0)}\;(n\to\infty) \right\}, 
\end{equation*}
where ``$\Rightarrow$" means the weak convergence. Moreover, 
\begin{equation*}
\mathcal{M}_W =\bigcup_{\Psi_0 \in \boldsymbol{\widetilde{\Phi}^{(o)}}}\mathcal{M}_W(\Psi_0). 
\end{equation*}
We call the element of $\mathcal{M}_W$ ``weak limit measure". 
We rewrite the element of $\mathcal{M}_W$ to emphasize ``weak convergence" and the dependence of the initial state by $\mu_W^{(\Psi_0)}$. 
For example, in the Hadamard walk case 
as already has been shown in \cite{Konno2002,Konno2005}, 
when the initial state is $\Psi_0=|0\rangle\otimes {}^T[1/\sqrt{2},i/\sqrt{2}]$, then 
the weak limit measure is 
$\mu^{(\Psi_0)}_W(dx)=f_K(x;1/\sqrt{2})dx$, where $f_K(x;r)$ is defined in Eq. (\ref{univ}). 
As is seen in \cite{IKS,FKK,MKK,SK,KM,LP}, typical weak limit measures are decomposed as follows: 
there exist a rational polynomial $w(y)$, real numbers $C\in [0,1)$ and $r\in(0,1)$ such that 
\begin{equation}\label{uni} \mu^{(\Psi_0)}(dx)=C\delta_0(x)+(1-C)w(x)f_K(x;r)dx. \end{equation}
We propose a universality class of QWs, $\mathcal{M}_{K} (\subseteq \mathcal{M}_W)$, 
characterized by the weak limit measure expressed as Eq. (\ref{uni}). 
That is, the measure can be written as the convex combination of the Dirac measure at the origin, $\delta_0(x)$, 
and the absolutely continuous part, $w(x)$ times $f_K(x;r)dx$. 
\section{Time averaged limit measure}
\subsection{Generating function for weight of passage}
In this subsection, we will consider a general setting such that each quantum coin is assigned to each position in the following way 
\[U_x=\begin{bmatrix} a_x & b_x \\ c_x & d_x \end{bmatrix}\in \mathrm{U}(2)  \]
with $a_xb_xc_xd_x\neq 0$, where $\mathrm{U}(2)$ is the set of $2\times 2$ unitary matrix. 
Let $\Xi(x,n)$ be the weight of all passage from the origin to the position $x$ at time $n$. For example, since 
$\Xi(1,3)$ is constructed by the weight of three passages ``left$\to$right$\to$right'', ``right$\to$left$\to$right'' and ``right$\to$right$\to$left'', so 
$\Xi(1,3)=Q_0Q_{-1}P_0+Q_0P_1Q_0+P_2Q_1Q_0$. 
For any weight of passages, 
there exist $p(x,n),q(x,n),r(x,n),s(x,n)\in \mathbb{C}$ such that $\Xi(x,n)=p(x,n)P_0+q(x,n)Q_0+r(x,n)R_0+s(x,n)S_0$, 
where $R_x=|L\rangle\langle R| U_x$, $S_x=|R\rangle\langle L| U_x$ and 
the product between them has an algebraic relation as shown in Table \ref{PQRS} (see \cite{Konno2002,Konno2005}), 
for example, $P_xR_0=a_xR_0$. 
\begin{table}
\caption{Relation between weights of moves }\label{PQRS}
\begin{center}
\begin{tabular}{c|cccc}
    & $P_0$ & $Q_0$ & $R_0$ & $S_0$ \\ \hline
$P_x$ & $a_x P_0$ & $b_xR_0$ & $a_xR_0$ & $b_xP_0$ \\ 
$Q_x$ & $c_xS_0$ & $d_xQ_0$ & $c_xQ_0$ & $d_xS_0$ \\
$R_x$ & $c_xS_0$ & $d_xR_0$ & $c_xR_0$ & $d_xP_0$ \\
$S_x$ & $a_xS_0$ & $b_xQ_0$ & $a_xQ_0$ & $b_xS_0$
\end{tabular}
\end{center}
\end{table}
At first, we give the generating function $\widetilde{\Xi}_x(z)\equiv\sum_t\Xi(x,n)z^t$ which is the key to solve the limit theorems treated here. 
Assume that we start the QW from the origin with the initial coin state $\psi_0={}^T[\alpha,\beta]$ throughout this section, 
where $|\alpha|^2+|\beta|^2=1$ and $T$ is the transposed operator. 
\begin{lem} \label{acco} Let $\Delta_x$ denote the determinant of $U_x$. Assume that $a_x$, $d_x\neq 0$ for all $x \in \mathbb{Z}$. 
 	\begin{enumerate}
	\item If $x=0$, then 
	\begin{equation}\label{pero0}
		\widetilde{\Xi}_0(z) 
		= \frac{1}{1-c_0\fp_0(z)-b_0\fm_0(z)-\Delta_0\fp_0(z)\fm_0(z)}
        		\begin{bmatrix} 1-b_0\fm_0(z) & d_0\fp_0(z) \\ a_0\fm_0(z) & 1-c_0\fp_0(z)\end{bmatrix}, 
	\end{equation}
        \item if $|x|\geq 1$, then
        \begin{equation} \label{pooh}\\                
		\widetilde{\Xi}_x(z)
		= \begin{cases}
                	\lp_{x-1}(z)\cdots\lp_{1}(z)
                 		\begin{bmatrix} \lp_{x}(z)\fp_x(z) \\ z\end{bmatrix}
                        	\begin{bmatrix} c_0 & d_0 \end{bmatrix}\widetilde{\Xi}_0(z) & \text{: $x\geq 1$, } \\
                                \\
                 	\lm_{x+1}(z)\cdots\lm_{-1}(z)
                 		\begin{bmatrix} z \\ \lm_{x}(z)\fm_x(z) \end{bmatrix}
                        	\begin{bmatrix} a_0 & b_0 \end{bmatrix}\widetilde{\Xi}_0(z) & \text{: $x\leq -1$, }       
        	  \end{cases}    
         \end{equation}
         \end{enumerate} 
where $\lp_x(z)=zd_x/(1-c_x\fp_x(z))$, $\lm_x(z)=za_x/(1-b_x\fm_x(z))$. 
Here $\widetilde{f}^{(\pm)}_x(z)$ has the following continued-fraction representation: 
\begin{align} 
	\fp_x(z) &= -\frac{z^2\Delta_{x+1}}{c_{x+1}}\left( 1-\frac{|a_{x+1}|^2}{1-c_{x+1}\fp_{x+1}(z)} \right),\label{pero+} \\
        \fm_x(z) &= -\frac{z^2\Delta_{x-1}}{b_{x-1}}\left( 1-\frac{|d_{x-1}|^2}{1-b_{x-1}\fm_{x-1}(z)} \right).\label{pero-} 
\end{align}
        

\end{lem}

\noindent {\it Proof}. 
Define $F^{(+)}(x,n)$ (resp. $F^{(-)}(x,n)$) as the weight of all passages which start from $x$ and return to the same position $x$ at time $n$ 
avoiding $\{y\in \mathbb{Z}: y\leq x\}$ (resp. $\{y\in \mathbb{Z}: y\geq x\}$) throughout the interval $0<s<n$, respectively. 
Furthermore denote $\Xi^{(+)}(x,n)$ (resp. $\Xi^{(-)}(x,n)$) as the weight of all passages which start from $x$ and return to the same position $x$ at time $n$ 
avoiding $\{y\in \mathbb{Z}: y< x\}$ (resp. $\{y\in \mathbb{Z}: y> x\}$) throughout the interval $0\leq s\leq n$. 
The generating function $\widetilde{F}^{(+)}_x(z)\equiv \sum_{n}F^{(+)}(x,n)z^n$ can be expressed as follows: 
there exists a complex number $\fp_x(z)$ such that $\widetilde{F}^{(+)}(z)=\fp_x(z)R_x$, where $R_x=|L\rangle\langle R|U(x)$. 
Remarking that $\widetilde{\Xi}^{(+)}_x(z)=I+\widetilde{F}^{(+)}_x(z)+\left(\widetilde{F}^{(+)}_x(z)\right)^2+\cdots=I+\widetilde{F}^{(+)}_x(z)\widetilde{\Xi}^{(+)}_x(z)$, 
we have 
	\begin{equation}\label{pero1}
	\widetilde{\Xi}^{(+)}_x(z)=\frac{1}{1-c_x\fp_x(z)}\begin{bmatrix} 1 & d_x\fp_x(z) \\ 0 & 1-c_x\fp_x(z) \end{bmatrix}. 
	\end{equation}
On the other hand, note that all the passages which generate the $F^{(+)}(x,n)$ exactly move to right at the first step and finally move to left. 
Then we have $\widetilde{F}^{(+)}_x(z)$ can be re-expressed as 
	\begin{equation}\label{pero2}
	\widetilde{F}^{(+)}_x(z)=zP_{x+1}\; \widetilde{\Xi}^{(+)}_{x+1}(z)\; zQ_x. 
	\end{equation}
Combining Eqs. (\ref{pero1}) and (\ref{pero2}), we obtain the continued-fraction representation of $\fp_x(z)$  given by Eq. (\ref{pero+}). 
In the same way, we obtain    
	\begin{equation}\label{pero1m}
	\widetilde{\Xi}^{(-)}_x(z)=\frac{1}{1-b_x\fm_x(z)}\begin{bmatrix} 1-b_x\fm_x(z) & 0 \\ a_x\fm_x(z) & 1 \end{bmatrix}, \;\;
	\widetilde{F}^{(-)}_x(z)=zQ_{x-1}\; \widetilde{\Xi}^{(-)}_{x-1}(z)\; zP_x. 
	\end{equation}
Thus the above equations induce the continued-fraction representation of $\fm_x(z)$  given by Eq. (\ref{pero-}). 
Next, let us consider the generating function $\widetilde{\Xi}_0(z)$. The relation between $\widetilde{\Xi}_0(z)$ and $\widetilde{F}_0^{(\pm)}(z)$,  
$\widetilde{\Xi}_0(z)=I+\left(\widetilde{F}_0^{(+)}(z)+\widetilde{F}_0^{(-)}(z)\right)+\left(\widetilde{F}_0^{(+)}(z)+\widetilde{F}_0^{(-)}(z)\right)^2+\cdots
=I+\left(\widetilde{F}_0^{(+)}(z)+\widetilde{F}_0^{(-)}(z)\right)\widetilde{\Xi}_0(z)$ implies Eq. (\ref{pero0}).  
Assume $x>0$. There exactly exists the time $0<s_*\leq n$ in the $n$-length passage from the origin to the position $x$ such that the passage never goes back to 
$\{y\in \mathbb{Z} : y<x\}$ for any time $s_*<s\leq n$. Therefore we get 
	\begin{equation}\label{pero3}
        \widetilde{\Xi}_x(z)=\widetilde{\Xi}_x^{(+)}(z)\;zQ_{x-1}\;\widetilde{\Xi}_{x-1}(z)   
	\end{equation}
Put $|u_y\rangle={}^T[\lp_y(z)\fp_y(z),\;\; z ], 
           |v_y\rangle={}^T[\overline{c_y}, \;\; \overline{d_y}]$, ($y\leq x$).                                     
Combining Eqs. (\ref{pero1}) and (\ref{pero3}), $\widetilde{\Xi}_x(z)$ can be re-written by 
        \begin{align*}
        \widetilde{\Xi}_x(z) &= |u_x\rangle\langle v_{x-1}|\widetilde{\Xi}_{x-1}(z)
        	              = \langle v_{x-1},u_{x-1} \rangle\cdots\langle v_1,u_1 \rangle |u_x \rangle\langle v_0| \widetilde{\Xi}_0(z) \\
                             &= \lp_{x-1}(z)\cdots\lp_{1}(z)
                 		\begin{bmatrix} \lp_{x}(z)\fp_x(z) \\ z\end{bmatrix}
                        	\begin{bmatrix} c_0 & d_0 \end{bmatrix}\widetilde{\Xi}_0(z).
        \end{align*}
On the other hand, for $x<0$, in a similar fashion, we obtain Eq. (\ref{pooh}). Thus we complete the proof. $\square$ \\
\noindent \\

We should remark that 
when $a_0=d_0=0$, $b_0=c_0=1$, $a_x=d_x=\rho$, $\bar{b}_x=-c_x=b$ with $\rho=\sqrt{1-|b|^2}$ for any $|x|\geq1$ and 
the initial state starts from the origin with the coin state $|L\rangle$, then the walk is nothing but the quantum walk on $\mathbb{Z}_+$ 
discussed in \cite{KS2010} of Type II QW. In this case, putting $\widetilde{f}^{(-)}_x(z)=0$ for all $x<0$, then we have 
\begin{equation}\label{Al_1} 
\widetilde{\Xi}_0(z)=\frac{1}{1-c_0\fp_0(z)}= \frac{2b}{2b-z\left\{ (z-z^{-1})+\sqrt{(z-z^{-1})^2+4|b|^2} \right\}}. 
\end{equation}
In the following, let us consider a connection between the generating function shown in Lemma 3.1 and the Carath{\'e}odory function related to the spectral measure
of the QW by confirming the correctness of Eq. (\ref{Al_1}) from another way in the view point of the spectral measure of the quantum walk \cite{CGMV}. 
Put $\mathcal{C}$ as the time evolution of Type II QW discussed in \cite{KS2010}. 
From the Maclaurin expansion and the Karlin-MacGregor formula \cite{KMcG} related to the QW \cite{CGMV}, 
the scalar valued Carath{\'e}odory function of the spectral measure $d\mu(z)$ of Type II QW, $F^{(II)}(z)$, is expressed as 
\begin{equation}\label{Al_3} 
F^{(II)}(z)\equiv \int_{|t|=1}\frac{t+z}{t-z}d\mu(t)= 1+2\sum_{j=1}^{\infty}\overline{(\mathcal{C}^j)}_{0,0}z^j, \;\;(|z|<1)
\end{equation} 
where $(\mathcal{C})_{0,0}$ is the element of left upper most of $\mathcal{C}$. 
Note that 
\begin{equation}\label{Al_4} 
(\mathcal{C}^j)_{0,0}=\Xi(0,j). 
\end{equation}  
Then combining Eq. (\ref{Al_3}) with Eq. (\ref{Al_4}), we get 
\begin{equation}\label{CaratheGene}
 (1+\overline{F^{(II)}(\bar{z})})/2=\widetilde{\Xi}_0(z). 
\end{equation} 
Indeed, by substituting the following explicit expression for $F^{(II)}(z)$ obtained in \cite{KS2010} by using the CGMV method into Eq. (\ref{CaratheGene}), 
we see that the same expression as RHS of Eq. (\ref{Al_1}) appears again. 
\begin{equation*}\label{CaratheII}
F^{(II)}(z)
=-\frac{(1-b)z-(1-\overline{b})z^{-1}}{bz+\overline{b}z^{-1}-\sqrt{(z-z^{-1})^2+4|b^2|}}.
\end{equation*}
\quad On the other hand, in the doubly infinite case,  
the matrix valued Carath{\'e}odory function $\boldsymbol{F}(z)$ following the order given by Eq. (1) in \cite{CGMV2} can be expressed by 
\begin{equation}\label{Al_5}
\boldsymbol{F}(z)=I+2\sum_{j=1}^{\infty} \overline{(\boldsymbol{\mathcal{C}}^j)_{\boldsymbol{0,0}}} z^j, 
\end{equation}
where 
\begin{equation} \label{Al_2}
(\boldsymbol{\mathcal{C}}^j)_{\boldsymbol{0,0}}
	=\begin{bmatrix}  \left({U^{(s)}}^{j}\right)_{0R,0R} &  \left({U^{(s)}}^j\right)_{0R,-1L} \\ 
                          \left({U^{(s)}}^j\right)_{-1L,0R} & \left({U^{(s)}}^j\right)_{-1L,-1L} \end{bmatrix}. 
\end{equation}
Here $\left({U^{(s)}}^j\right)_{mJ,m'J'}=\langle m,J|{U^{(s)}}^j|m',J' \rangle$, ($m,m'\in\mathbb{Z}$, $J,J'\in \{ L,R \}$). 
Let $\sigma(\widetilde{\Xi}_x(z))$ be the generating function of the weight of passages from the origin to $x$ 
in the setting where every quantum coin $U_y$ is assigned to $y+1$ ($y\in \mathbb{Z}$). 
From a similar argument, 
a relation between our generating functions and the matrix valued Carath{\'e}odory function corresponding to Eq. (\ref{CaratheGene}) is 
given by 
\begin{equation}\label{CaratheGene2}
\frac{1}{2}\left(I+\overline{\boldsymbol{F}(\bar{z})}\right)
=\begin{bmatrix} \langle R|\widetilde{\Xi}_0(z)|R\rangle & \langle R|\sigma(\widetilde{\Xi}_1(z))|L\rangle \\
                 \langle L|\widetilde{\Xi}_{-1}(z)|R\rangle & \langle L|\sigma(\widetilde{\Xi}_0(z))|L\rangle
 \end{bmatrix}.
\end{equation}
Therefore a relation between our generating function and the Carath{\'e}odory fucntion is given by 
Eq. (\ref{CaratheGene}) (semi-infinite case) and Eq. (\ref{CaratheGene2}) (doubly infinite case). 
Another expression for the Carath{\'e}odory function $\boldsymbol{F}(z)$ using the Schur functions can be seen in \cite{CGMV2}. 
The asymptotic return probability and its necessary and sufficient condition for the localization of one defect QW 
are explicitly obtained with nice geometric representations in \cite{CGMV2} by using the CGMV method. 
\subsection{Time averaged limit measure of QW with one defect}
In this subsection, we consider a quantum walk with one defect at the origin including \cite{KonnoLocal}, that is, 
\begin{equation}\label{setting}
U_x= \begin{cases}
      U_0=\begin{bmatrix} a_0 & b_0 \\ c_0 & d_0 \end{bmatrix} & \text{: $x=0$,} \\
      \\
        U=\begin{bmatrix} a & b \\ c & d \end{bmatrix} & \text{: $|x|\geq 1$}
     \end{cases}
\end{equation}
with $\Delta=\Delta_0$, where 
$\Delta$ and $\Delta_0$ are the determinants of $U$ and $U_0$, respectively. 
From Lemma \ref{acco}, we obtain the following time averaged limit measure. Furthermore, 
the weak limit measure is given in the next section. \\

\noindent \\
\noindent \\
\begin{thm}\label{acco1} 
Let the QW with one defect be launched at the origin with the initial coin state $\varphi_0={}^T[\alpha,\beta]$ with $|\alpha|^2+|\beta|^2=1$. 
Put $m=\mathrm{Re}(\bar{c}c_0)$. Then we have the following time averaged limit measure: 
\begin{enumerate}
\item if $x=0$,
\begin{equation}
\overline{\mu}_\infty^{(\Psi_0)}(0)
	=I_{\{|c|^2>m\}}\frac{1}{2}\left(\frac{2(|c|^2-m)}{1-2m+|c|^2}\right)^2, 
\end{equation}
\item if $|x|\geq 1$, 
\begin{multline}
\overline{\mu}_\infty^{(\Psi_0)}(x)
	=\overline{\mu}_\infty^{(\Psi_0)}(0)\frac{|c|^2(1-m)}{(|c|^2-m^2)(1-2m+|c|^2)}\left(\frac{|a|^2}{1-2m+|c|^2}\right)^{|x|-1} \\
        \times 
        \begin{cases}
        \left(1+|c_0|^2-2m^2/|c|^2\right)|\alpha|^2+|a_0|^2|\beta|^2+2\mathrm{Re}\left(\alpha\bar{\beta}\bar{d_0}(c_0-m/\bar{c})\right) & \text{: $x\geq 1$,} \\
        \left(1+|c_0|^2-2m^2/|c|^2\right)|\beta|^2+|a_0|^2|\alpha|^2-2\mathrm{Re}\left(\alpha\bar{\beta}\bar{d_0}(c_0-m/\bar{c})\right) & \text{: $x\leq -1$.}
        \end{cases}        
\end{multline}
\end{enumerate}
\end{thm}
\noindent {\it Proof.} 
Remark that $\fp_x(z)=\fp$, $\lp_x(z)=\lp$ for all $x\geq 0$ and $\fm_x(z)=\fm(z)$, $\lm_x(z)=\lm(z)$ for all $x\leq 0$, 
where $\lp=zd/(1-c\fp)$, $\lm=za/(1-a\fm)$. 
Here $\fp$ and $\fm$ are one of the solutions for 
	\[(\fp)^2+w(w-w^{-1})/c\;\fp-w^2(|c|/c)^2=0 \;\; \mathrm{and}\; (\fm)^2+w(w-w^{-1})/c\;\fm-w^2(|b|/b)^2=0 \] 
which derive from Eqs. (\ref{pero+}) and (\ref{pero-}), respectively, where 
\begin{equation}\label{john1}
w=\Delta^{1/2} z. 
\end{equation}
Thus the explicit expressions for $\widetilde{\theta}^{(\pm)}$ and $\widetilde{\lambda}^{(\pm)}$ are 
\[ \fp=-\frac{w}{2c}\left\{ (w-w^{-1})+\sqrt{(w-w^{-1})^2+4|c|^2} \right\}, \;\;
   \fm=-\frac{w}{2b}\left\{ (w-w^{-1})+\sqrt{(w-w^{-1})^2+4|b|^2} \right\},\]
\[   \lp=\frac{\Delta^{1/2}}{2a}\left\{ (w+w^{-1})-\sqrt{(w+w^{-1})^2-4|a|^2} \right\},\;\;
   \lm=\frac{\Delta^{1/2}}{2d}\left\{ (w+w^{-1})-\sqrt{(w+w^{-1})^2-4|d|^2} \right\}.  \]
We choose the square root so that $|\widetilde{\lambda}^{(\pm)}|<1$ for $|z|<1$ and for $|z|\to 0$, $\widetilde{f}^{(\pm)}\to 0$. 
If $w=(1-\epsilon)e^{i\theta}$, then the square root is expressed as \cite{CGMV2}
\begin{equation}\label{CGMV}
	\lim_{\epsilon\downarrow 0}\sqrt{(w+w^{-1})^2-4|a|^2}=
        	\begin{cases}
		2\;\mathrm{sgn}(\cos\theta)\sqrt{|c|^2-\sin^2\theta} & \text{: $|a|\leq|\cos\theta|$, } \\
        	-2i\;\mathrm{sgn}(\sin\theta)\sqrt{|a|^2-\cos^2\theta} & \text{: $|a|>|\cos\theta|$. } 
		\end{cases}
\end{equation}
Put $m=\mathrm{Re}(\overline{c}c_0)$ and 
$\widetilde{\Lambda}_0=1-c_0\fp_0(z)-b_0\fm_0(z)-\Delta_0\fp_0(z)\fm_0(z)$. 
Each of four poles of $\widetilde{\Xi}_x(z)$ with the absolute values $1$ 
is equal to an each of solutions of $\widetilde{\Lambda}_0=0$ given by 
\begin{equation}\label{john2}
\pm w_{\pm}= \pm (1+|c|e^{\pm i\gamma})/|1+|c|e^{\pm i\gamma}|, 
\end{equation}
where $\cos\gamma =-m/|c|$. 
Noting $|a|<|\mathrm{Re}(\pm w_{\pm})|$, Eq. (\ref{CGMV}) implies 
	\begin{align} 
        	\lim_{w\to \pm \omega_+} \frac{w\pm \omega_+}{\widetilde{\Lambda}_0} &= \mp I_{\{|c|^2>m\}}\frac{ie^{-i\gamma}\omega_+}{2}\frac{|c|}{\sqrt{|c|^2-m^2}}
		\frac{|c|^2-m}{1-2m+|c|^2}, \label{pooh1} \\  
		\lim_{w\to \pm \omega_-} \frac{w\pm \omega_-}{\widetilde{\Lambda}_0} &=\pm I_{\{|c|^2>m\}}\frac{ie^{-i\gamma}\omega_-}{2}\frac{|c|}{\sqrt{|c|^2-m^2}}
		\frac{|c|^2-m}{1-2m+|c|^2}. \label{pooh2}
        \end{align}  
From the Cauchy theorem and the description of pp.264-265 in \cite{Flajolet}, 
the asymptotic behavior of $\Xi(x,n)$ for large time step $n$ can be evaluated as 
\begin{align} 
	\Xi(x,n) &= \frac{1}{2\pi i}\int_{|z|=r_0} \widetilde{\Xi}_x(z)\frac{dz}{z^{n+1}} \label{pooh3}\\
        	 &= M^{(+)}_+\omega_+^{-(n+1)}+M^{(-)}_+(-\omega_+)^{-(n+1)}+M^{(+)}_-\omega_-^{-(n+1)}+M^{(-)}_-(-\omega_-)^{-(n+1)}+O(n^{-1}), \label{pooh4}       
\end{align}
where $M^{(\pm)}_\epsilon=-\mathrm{Res}(\widetilde{\Xi}_x(z); \pm w_{\epsilon})$, $\epsilon\in \{\pm\}$. 
Therefore from Proposition \ref{acco}, Eqs. (\ref{pooh1})-(\ref{pooh4}) provide 
\begin{multline} \label{hinano}
	\Xi(x,n)\sim 
                \frac{1+(-1)^{x+n}}{2}\times I_{\{|c|^2>m\}}\times \frac{-i|c|(|c|^2-m)}{(1-2m+|c|^2)\sqrt{|c|^2-m^2}}\\ 
                \times \begin{cases}
                	\left\{(e^{-i\gamma}I+|c|M_0)\omega_+^{-n}-(e^{i\gamma}I+|c|M_0)\omega_-^{-n}\right\} & \text{: $x=0$, } \\
                        \\
                	r_+^{x-1}
                        \left( \begin{bmatrix} -r_+|c|/c \\ \Delta^{-1/2}\omega_+ \end{bmatrix}
                               \begin{bmatrix} c_0+e^{i\gamma}c/|c| & d_0 \end{bmatrix} \omega_+^{-n}
                               -\begin{bmatrix} -r_+|c|/c \\ \Delta^{-1/2}\omega_- \end{bmatrix}
                                \begin{bmatrix} c_0+e^{-i\gamma}c/|c| & d_0 \end{bmatrix} \omega_-^{-n}
                        \right) & \text{: $x\geq 1$,} \\
                        \\
                        r_-^{|x|-1}
                        \left( \begin{bmatrix} \Delta^{-1/2}\omega_+ \\  -r_-|b|/b \end{bmatrix}
                               \begin{bmatrix} a_0 & b_0+e^{i\gamma}b/|b| \end{bmatrix} \omega_+^{-n}
                               -\begin{bmatrix} \Delta^{-1/2}\omega_- \\ -r_+|b|/b \end{bmatrix}
                                \begin{bmatrix} a_0 & b_0+e^{-i\gamma}b/|b| \end{bmatrix} \omega_-^{-n}
                        \right) & \text{: $x\leq -1$,} 
                       \end{cases} 
\end{multline} 
where $r_+=d\Delta^{-1/2}/\sqrt{1-2m+|c|^2}$, $r_-=a\Delta^{-1/2}/\sqrt{1-2m+|b|^2}$ and 
\[ M_0=\begin{bmatrix} b_0/b & -d_0/c \\ -a_0/b & c_0/c \end{bmatrix}.  \]
Thus since $P(X_n=x)=||\Xi(x,n)\psi_0\rangle||^2$ and $\lim_{T\to\infty}\frac{1}{T}\sum_{n=0}^{T-1}(\omega_\epsilon/\omega_\tau)^n=\delta_{\epsilon,\tau}$ 
($\epsilon,\tau\in\{\pm\}$), 
we obtain the desired conclusion. 
\begin{flushright} $\square$ \end{flushright}

\begin{cor}{\rm (In the setting of \cite{KonnoLocal} case)} 
Let the initial state $\Psi_0$ start from the origin with the coin state $\psi_0={}^T[1/\sqrt{2},i/\sqrt{2}]$.  
Assume $U=U(0,0)$ and $U_0=U_0(0,\omega)$ given by Eq. (\ref{seikou}). 
Then we have 
\begin{enumerate}
	\item if $x=0$, 
	\begin{equation}
        \overline{\mu}_\infty^{(\Psi_0)}(0)=2I_{\{\omega \notin 2n\pi: n\in\mathbb{Z}\}}(\omega)\left(\frac{1-\cos \omega}{3-2\cos \omega}\right)^2,
        \end{equation}
        \item if $|x|\geq 1$, 
        \begin{equation}
        \overline{\mu}_\infty^{(\Psi_0)}(x)=\overline{\mu}_\infty^{(\Psi_0)}(0)\frac{2-\cos\omega}{(3-2\cos\omega)^{|x|}}
        	\times \begin{cases}
                	\left(1+\frac{\sin\omega}{1+\sin^2\omega}\right) & \text{: $x\geq 1$, } \\
                        \\
                        \left(1-\frac{\sin\omega}{1+\sin^2\omega}\right) & \text{: $x\leq -1$. }        
                       \end{cases}
        \end{equation}
\end{enumerate}
\end{cor}

\begin{rem} 
{\rm
The localization emerges in the QW with one defect for the case of $\Delta=\Delta_0$, if and only if $|c|^2>m$ depending on the initial coin state. 
The condition agrees with ($M^4$) in their notation of \cite{CGMV2}. We also confirm that the localization dose not occurs 
without defect because of $|c|^2=m$. }
\end{rem}
\begin{rem}
{\rm 
Let $\boldsymbol{1}$ be the all one vector on $\mathbb{C}^{\mathbb{Z}}$. Then we have 
\[ \langle \boldsymbol{1}, \overline{\mu}_{\infty}^{(\Psi_0)} \rangle < 1,\]
Therefore we have $\overline{\mu}_{\infty}^{(\Psi_0)}\in \overline{\mathcal{M}}_\infty \setminus \mathcal{M}_I$. }
\end{rem}
We discuss the missing value ``$1-\langle \boldsymbol{1}, \overline{\mu}_{\infty} \rangle$" in the next section. 
\section{Weak limit measure of QW with one defect}
Let $C$ be the summation of the time averaged limit measure $\overline{\mu}_\infty^{(\Psi_0)}(x)$ 
obtained by Theorem \ref{acco1} over all the positions $x\in \mathbb{Z}$. 
We should remark $0\leq C<1$. The following is the limit theorem with respect to the missing value $1-C$: 
\begin{thm} \label{acco3} 
Let the QW with one defect in the choice of quantum coins given by Eq. (\ref{setting}) 
start from the origin with the initial coin state $\varphi_0={}^T[\alpha,\beta]$. 
Then it is obtained that as $n\to \infty$,         
$X_n/n$ weakly converges to $Z$ which has the following density function $\rho(x)$
\[\rho(x)=C\delta(x)+ w(x)f_K(x;|a|), \]
where $f_K(x;|a|)$ is defined in Eq. (\ref{univ}) and 
\[w(x)= \frac{|c|^2x^2}{(|c|^2-m)^2+(|c|^2-m^2)x^2}\left[ \gamma(x)-|a_0|^2\left\{ (|\alpha|^2-|\beta|^2)
	+\frac{2\mathrm{Re}(a\alpha\overline{b\beta})}{|a_0|^2} \right\}x \right]. 
 \]
Here 
\[
\gamma(x)
=\begin{cases}
|\alpha|^2(1-2m+|c_0|^2)+|\beta|^2|a_0|^2+2\mathrm{Re}\left(a_0\alpha\overline{(b-b_0)\beta}\right) & \text{: $x\geq 0$,} \\
|\beta|^2(1-2m+|c_0|^2)+|\alpha|^2|a_0|^2-2\mathrm{Re}\left(a_0\alpha\overline{(b-b_0)\beta}\right) & \text{: $x< 0$.}
\end{cases}
\]
\end{thm}
\begin{rem}
{\rm The QW with one defect also belongs to $\mathcal{M}_{K}$ 
as in the case of typical homogeneous QWs \cite{IKS,FKK,MKK,SK,KM,LP,CHKS,CKS}.} 
\end{rem}
\noindent \\
\noindent {\it Proof. } 
We have already got the generating function of $\Xi(x,t)$ in Proposition \ref{acco}. 
Moreover, we consider its Fourier transform $\widehat{\widetilde{\Xi}}(k,z)\equiv \sum_{x\in \mathbb{Z}}\widetilde{\Xi}_x(z)e^{ikx}$. 
Since there exist matrix valued constants $C_+$, $C_0$, $C_-$ which are independent of position $x$ such that 
$\widetilde{\Xi}_x(z)=C_+\widetilde{\lambda}_+^{x}+C_0+C_-\widetilde{\lambda}_-^{|x|}$ with $|\widetilde{\lambda}^{(\pm)}|<1$ for $|z|<1$, 
an explicit expression for $\widehat{\widetilde{\Xi}}(k,z)$ is 
\begin{multline}
	\widehat{\widetilde{\Xi}}(k,z)
        	=\frac{1}{\widetilde{\Lambda}_0}\bigg\{
                (I-c\fp M_0)+\frac{e^{ik}}{1-e^{ik}\lp}\begin{bmatrix} \fp\lp \\ z \end{bmatrix} \begin{bmatrix} c_0+\Delta_0\fm & d_0 \end{bmatrix} \\
                +\frac{e^{ik}}{1-e^{-ik}\lm}\begin{bmatrix} z \\ \fm\lm \end{bmatrix} \begin{bmatrix} a_0 & b_0+\Delta_0\fp \end{bmatrix} \bigg\},\;\;(|z|<1)
\end{multline}
The poles deriving from $\widetilde{\Lambda}_0=0$ correspond to localization, i.e., the Dirac measure $C\delta_0(x)$. 
On the other hand, the important poles which give a absolutely continuous part of the universality class come from 
\begin{equation}\label{universal}
	1-e^{\pm ik}\widetilde{\lambda}^{(\pm)}=0. 
\end{equation}
Noting Eq. (\ref{CGMV}), the solutions for Eq.(\ref{universal}) are $e^{\pm i\varphi(s)}$, where $\cos \varphi(s)=|a|\cos s$ 
and $\mathrm{sgn}(\sin\varphi(s))=-\mathrm{sgn}(\sin s)$ with $s=k+\mathrm{arg}(\overline{a}\Delta^{1/2})$, respectively. 
Let $\widehat{\Xi}_n(k)=\sum_{x\in \mathbb{Z}} \Xi(x,n)e^{ikx}$ and $M_n(k,\xi)=\widehat{\Xi}_n^{\dagger}(k)\widehat{\Xi}_n(k+\xi)$. 
Since $||\widehat{\widetilde{\Xi}}(k,z)||<\infty$ for $|z|<1$,  from the Fubini theorem, $\widehat{\widetilde{\Xi}}(k,z)$ can be reexpressed by 
$\widehat{\widetilde{\Xi}}(k,z)=\sum_{n\geq 0}\widehat{\Xi}_n(k)z^n$ for $|z|<1$. 
The Cauchy theorem and \cite{Flajolet} provide the asymptotics of $\widetilde{\Xi}_n(k)$ in large $n$ as follows: 
\begin{equation}
	\widehat{\Xi}_n(k)\sim e^{-i(n+1)\varphi(s)}V_+(s)+e^{i(n+1)\varphi(s)}V_-(s)+L(s) 
\end{equation}
where $V_{\pm}(s)$ are the matrix valued residues at the unit pole $e^{\pm i\varphi(s)}$, 
i.e., $V_{\pm}(s)=-\mathrm{Res}(\widehat{\widetilde{\Xi}}(k,z); e^{\pm i\varphi(s)})$, respectively, and 
\[ L(s)=\sum_{j\in \{0,1\},\epsilon \in \{\pm\} }\left( (-1)^j w_{\epsilon} \right)^{n+1} L_{j,\epsilon}(s),\] 
Here $L_{j,\epsilon}(s)$ is the matrix valued residues at the unit pole $(-1)^j w_\epsilon$. 
We should remark that the characteristic function for the QW $X_n$ with the initial coin state $\psi_0$ is obtained by 
\begin{equation}
	E[e^{i\xi X_n}]=\int_{-\pi}^{\pi}  \langle \psi_0, M_n(k,\xi)\psi_0\rangle \frac{dk}{2\pi}. 
\end{equation}
Because the behavior has a ballistic spreading, from now on, we will concentrate on the evaluation for $M_n(k,\xi/n)$. 
Let 
	\[ h(s)\equiv \partial\varphi(s)/\partial s=-|a|\frac{|\sin s|}{\sqrt{1-|a|^2\cos^2s}}. \]
Then we have from the Riemann-Lebesgue lemma, for large $n$, 
\begin{equation}\label{gosaku}
	E[e^{i\xi X_n/n}]
        	\sim C
                	+\int_{-\pi}^{\pi}  e^{-ih(s)}\langle \psi_0,P_+(s)\psi_0\rangle\frac{dk}{2\pi}
                	+\int_{-\pi}^{\pi}  e^{ih(s)}\langle\psi_0,P_-(s)\psi_0\rangle \frac{dk}{2\pi},
\end{equation}
where $P_\pm(s)=V_\pm^\dagger(s) V_\pm(s)$ and 
\[ C=\int_{-\pi}^{\pi} ||L(s)|\psi_0\rangle ||^2 \frac{ds}{2\pi}=\sum_{x\in \mathbb{Z}}\overline{\mu}_\infty(x).  \]
Notice that if $w=e^{i\pm \varphi(s)}$, then $|\lambda^{(\pm)}|=1$. The explicit expression for $P_{\pm}(s)$ are 
\begin{align}
	P_+(s) &= \frac{\widetilde{m}_+(1+|\fp_+|^2)}{|\widetilde{\Lambda}_{0,+}|^2}
        		\begin{bmatrix} |c_0+\Delta_0\fm_+|^2 & \overline{(c_0+\Delta_0\fm_+)}d_0 \\ (c_0+\Delta_0\fm_+)\overline{d}_0 & |d_0|^2 \end{bmatrix}, \\
        P_-(s) &= \frac{\widetilde{m}_-(1+|\fm_-|^2)}{|\widetilde{\Lambda}_{0,-}|^2}
        	  	\begin{bmatrix} |a_0|^2  & (b_0+\Delta_0\fp_-)\overline{a}_0 \\ \overline{(b_0+\Delta_0\fp_-)}a_0 & |b_0+\Delta_0\fp_-|^2 \end{bmatrix},              
\end{align}
where $\widetilde{m}_{\pm}=|\mathrm{Res}(1/(1-e^{\pm ik}\widetilde{\lambda}^{(\pm)}); e^{\pm i\varphi(s)})|^2$, 
$\widetilde{f}^{(\epsilon)}_{\pm}=\lim_{w\to e^{\pm i\varphi(s)}}\widetilde{f}^{(\epsilon)}(w)$, ($\epsilon\in\{\pm\}$), and 
$\widetilde{\Lambda}_{0,\pm}=\lim_{w\to e^{\pm i\varphi(s)}}\widetilde{\Lambda}_0(w)$. 
If we substitute $x=h(s)$ to the integration in Eq. (\ref{gosaku}), then the following relations hold: 
\begin{align}
\cos s &= \frac{\mathrm{sgn}(\cos s)}{|a|}\sqrt{\frac{|a|^2-x^2}{1-x^2}},\;\;\sin s = \mathrm{sgn}(\sin s)\frac{|c|}{|a|}\frac{|x|}{\sqrt{1-x^2}}, \label{tokuzo1} \\
ds &= -\mathrm{sgn}(\sin 2s)\pi f_K(x;a)dx. 
\end{align}
Equation (\ref{tokuzo1}) gives $\widetilde{m}_{\pm}=x^2$, $1+|\fp_{+}|^2=1+|\fm_{-}|^2=2/(1+x)$, 
and 
\begin{align*}
\Delta_0\fm_{+} &= \frac{-c/|c|\left(|c|+i\;\mathrm{sgn}(\sin 2s)\sqrt{|a|^2-x^2}\right)}{1+x}, \;\;\;
\Delta_0\fp_{-} &= \frac{-b/|b|\left(|b|-i\;\mathrm{sgn}(\sin 2s)\sqrt{|d|^2-x^2}\right)}{1+x}, \\
\end{align*}
and finally 
\[ \frac{1}{|\widetilde{\Lambda}_{0,\pm}|^2} = \frac{|c|^2(1+x)^2}{4\{(|c|^2-m)^2+(|c|^2-m^2)x^2\}}.\]
Substituting these values into Eq. (\ref{gosaku}) completes the proof. \begin{flushright}$\square$\end{flushright}

\begin{rem}
{\rm
It can be seen that the density function always takes zero at the origin except $m=|c|^2$ case. 
Moreover in the no defect case, the density function with the initial coin state ${}^T[\alpha,\beta]$ is written as 
	\[ \rho(x)=\left\{1-\left(|\alpha|^2-|\beta|^2+2\mathrm{Re}(a\alpha\overline{b\beta})/|a|^2\right)x \right\}f_K(x;a) \]
which agrees with \cite{Konno2002,Konno2005}.  }       
\end{rem}
\begin{cor}{\rm (In the setting of \cite{KonnoLocal})} 
Let the initial state $\Psi_0$ start from the origin with the coin state $\psi_0={}^T[1/\sqrt{2},i/\sqrt{2}]$.  
Assume $U=U(0,0)$ and $U_0=U_0(0,\omega)$ given by Eq. (\ref{seikou}). 
Then we have 
\begin{equation*}
	X_n/n\Rightarrow Z \;\;\;(n\to\infty),
\end{equation*}
where $Z$ has the following density $\rho(x)$: 
\begin{equation*}
   \rho(x)=C\delta_0(x)
        +w(x)f_K(x;1/\sqrt{2}),
\end{equation*}
where 
\begin{align*} 
	C	=2\left(\frac{1-\cos\omega}{3-2\cos\omega}\right)^2, \;\;
        w(x)	= \frac{2-\cos\omega+\mathrm{sgn}(x)\sin\omega+1/2\cdot x \sin\omega}{(1-\cos\omega)^2+(2-\cos^2\omega)x^2}x^2. 
\end{align*} 
\end{cor}
\section{Discussions}
In this section, for simplicity, we restrict ourselves to treat the quantum coins in the setting of \cite{KonnoLocal}, \textit{i.e.,} 
$U$ is the Hadamard matrix and 
\[ U_0=\frac{1}{\sqrt{2}}\begin{bmatrix} 1 & e^{i\omega} \\ e^{-i\omega} & -1 \end{bmatrix}. \]
Let the eigenvalue of the time evolution of the one defect QW 
$U^{(s)}$ be $\eta$, and corresponding eigenvector be 
\[ \Psi(\eta) = {}^T \left[ \ldots, \Psi^{(L)}(-1), \Psi^{(R)} (-1), \Psi^{(L)} (0), \Psi^{(R)} (0), \Psi^{(L)} (1), \Psi^{(R)} (1), \ldots \right], \]
that is $U^{(s)}\Psi(\eta)=\eta\Psi(\eta)$. 
Let 
$\widehat{\Psi}_+^{(L)} (z)=\sum_{x\geq 1}\Psi^{(L)} (x)z^x$, 
$\widehat{\Psi}_+^{(R)} (z)=\sum_{x\geq 1}\Psi^{(R)} (x)z^x$, 
$\widehat{\Psi}_-^{(L)} (z)=\sum_{x\leq -1}\Psi^{(L)} (x)z^x$ and 
$\widehat{\Psi}_-^{(R)} (z)=\sum_{x\leq -1}\Psi^{(R)} (x)z^x$. 
A directly computation gives the generating functions for $\Psi^{(L)}(x)$, $\Psi^{(R)}(x)$ as follows: 
there exist $C_j^{(J,\pm)}(\eta)\in \mathbb{C}$ which are independent of the parameter $z$ ($j\in \{1,2\}$, $J\in{L,R}$) such that 
\begin{align*} 
\widehat{\Psi}_+^{(L)} (z) &= \frac{C_1^{(L,+)}(\eta)}{1+\gamma z}+\frac{C_2^{(L,+)}(\eta)}{1-z/\gamma}-\Psi^{(L)}(0), \;
\widehat{\Psi}_-^{(L)} (z) = \frac{C_1^{(L,-)}(\eta)}{1+\gamma z}+\frac{C_2^{(L,-)}(\eta)}{1-z/\gamma}, \\
\widehat{\Psi}_+^{(R)} (z) &= \frac{C_1^{(R,+)}(\eta)}{1+\gamma z}+\frac{C_2^{(R,+)}(\eta)}{1-z/\gamma}, \;
\widehat{\Psi}_-^{(R)} (z) = \frac{C_1^{(R,-)}(\eta)}{1+\gamma z}+\frac{C_2^{(R,-)}(\eta)}{1-z/\gamma}-\Psi^{(R)}(0), \\
\end{align*}
where $\gamma$ is a solution for 
\begin{equation}
h(z) = z^2 + \sqrt{2} \left( \eta - \frac{1}{\eta} \right) z  - 1 
\end{equation}
with $|\gamma|<1$. 
Here $C_1^{(L,+)}=\Psi^{(L)}(0)$, $C_2^{(L,-)}=\Psi^{(R)}(0)/(\sqrt{2}\eta\gamma-1)$, 
$C_1^{(R,+)}=-\Psi^{(L)}(0)/(\sqrt{2}\eta\gamma-1)$ and $C_2^{(R,-)}=\Psi^{(R)}(0)$. 
A heuristic argument gives 
\begin{equation}\label{heuristic} 
C_2^{(L,+)}(\eta)=C_1^{(L,-)}(\eta)=C_2^{(R,+)}(\eta)=C_1^{(R,-)}(\eta)=0. 
\end{equation}
Indeed, we can confirm that $U^{(s)}\Psi(\eta)=\eta\Psi(\eta)$ under Eq. (\ref{heuristic}). 
Thus we have an expression for the eigenvector whose eigenvalue is $\eta$ with two parameters $\Psi^{(L)}(0)$, $\Psi^{(R)}(0)\in \mathbb{C}$: 
\begin{equation}\label{john10}
\Psi^{(L)}(j) = \begin{cases} 
C_1^{(L,+)}(-\gamma)^j & \text{: $j\geq 1 $, } \\ 
\Psi^{(L)}(0) & \text{: $j=0$,} \\ 
C_2^{(L,-)}\gamma^{|j|} & \text{: $j\leq -1$,} 
\end{cases}
\end{equation}
\begin{equation}\label{john11}
\Psi^{(R)}(j) = \begin{cases} 
C_1^{(R,+)}(-\gamma)^j & \text{: $j\geq 1 $, } \\ 
\Psi^{(R)}(0) & \text{: $j=0$,} \\ 
C_2^{(R,-)}\gamma^{|j|} & \text{: $j\leq -1$.} 
\end{cases}
\end{equation}
Combining Eq. (\ref{john10}) for $j=-1$ and Eq. (\ref{john11}) for $j=1$ with the definition of $\Psi(\eta)$ provides the following 
explicit expression for the eigenvalue $\eta$: 
\[ \eta\in \mathcal{E}\equiv \left\{\frac{\sigma \sqrt{2-\cos^2 \omega}+\tau i (2-\cos \omega)}{\sqrt{2}\sqrt{3-2\cos\omega}}: \sigma, \tau\in \{\pm 1\}\right\}. \]
\begin{rem}
Put $\mathcal{S}$ as the set of four singular points of the generating function of weight of passages given by Eq. (\ref{john2}) as follows.
\[ \mathcal{S}=\{\sigma w_{\tau}: \sigma,\tau \in\{\pm \}\}. \] 
Denote $\mathcal{M}$ as the set of four mass points of the spectral measure of $U^{(s)}$ obtained by \cite{CGMV2} which belongs to 
the class $(M^4)$ in the notation of \cite{CGMV2}. Then we find that
\[ \mathcal{E}=i\mathcal{S}=\mathcal{M}, \]
where $i\mathcal{S}=\{iw: w\in \mathcal{S}\}$. The phase $i$ in the above equation comes from the transformation given by Eq. (\ref{john1}). 
(in this case, $\Delta=-1$.) 
\end{rem}
We have concentrated on the initial condition, $\Psi_0$, starting from one position. 
From now on, using the above arguments, 
we discuss a stationary measure when we remove the restriction to the initial condition. 
\begin{pro}\label{dis}
Assume that the walk starts with an initial state $\Psi_0 \in \{\widetilde{\Psi}_0^{(\sigma,\tau)}: \sigma,\tau\in\{\pm 1\}\}$, where
\begin{align*} 
\widetilde{\Psi}_0^{(\sigma,\tau)}(x)=
	\left(\frac{\tau i\mathrm{sgn}(x)}{\sqrt{3-2\cos\omega}}\right)^{|x|}\times 
        	\begin{cases} 
                \phi_0^{(L)}\times \begin{bmatrix} 1 \\ (1-\cos\omega)-\sigma\tau i\sqrt{2-\cos^2\omega} \end{bmatrix} & \text{: $x\geq 1$, } \\ 
                \\
                \begin{bmatrix} \phi_0^{(L)} \\ \phi_0^{(R)} \end{bmatrix} & \text{: $x=0$,}\\ 
                \\ 
                \phi_0^{(R)}\times \begin{bmatrix} -(1-\cos\omega)+\sigma\tau i\sqrt{2-\cos^2\omega} \\ 1 \end{bmatrix} & \text{: $x\leq -1$. } 
                \end{cases} 
\end{align*} 
Then the stationary measure with the initial state $\widetilde{\Psi}_0^{(\sigma,\tau)}$, $\mu_I^{(\widetilde{\Psi}_0^{(\sigma,\tau)})}$, is given by 
\begin{align*} 
\mu_I^{(\widetilde{\Psi}_0^{(\sigma,\tau)})}(x)=
	\left( \frac{1}{3-2\cos\omega} \right)^{|x|}\times 
		\begin{cases} 2(2-\cos\omega)|\phi_0^{(L)}|^2 & \text{: $x\geq 1$,} \\ 
		|\phi_0^{(L)}|^2+|\phi_0^{(R)}|^2 & \text{: $x=0$,} \\ 
		2(2-\cos\omega)|\phi_0^{(R)}|^2 & \text{: $x\leq -1$.} 
		\end{cases} 
\end{align*}
\end{pro}
We should remark that when $\omega=0$ and we put the two parameters in Proposition \ref{dis} as $|\phi^{(L)}_0|^2=|\phi^{(R)}_0|^2=c/2$ ($c>0$), 
then we obtain the eigenvector $\Psi$ satisfying 
$|\Psi^{(R)}(x)|^2=|\Psi^{(L)}(x)|^2=c/2$ for all $x\in \mathbb{Z}$. 
Therefore we get the uniform measure $\mu_I^{(\widetilde{\Psi}_0^{(\sigma,\tau)})}(x)=c$ for all $x \in \mathbb{Z}$ as 
a stationary measure of the Hadamard walk. \\
Let us also remark that by selecting an appropriate
initial conditions, one can avoid localizations
even for the systems with one defect for which the localization always exists 
when we start with an initial state which has all the support at the point of the defect. 
Indeed, for that, it is enough to compute one of at most four eigenvectors of $U^{(s)}$ and 
choose an initial state which is orthogonal to all of them. \\

Let $\Psi_0$ be the initial state with $\Psi_0(0)={}^T[1/\sqrt{2},i/\sqrt{2}]$ and $\Psi_0(x)={}^T[0,0]$ $(x\neq 0)$. 
From Eq. (\ref{hinano}), we obtain time averaged limit measure of left and right chirarities at the origin: 
\begin{align*}
\overline{\mu}_\infty^{(\Psi_0,L)}(0)\equiv \lim_{n\to \infty}\frac{1}{T}\sum_{n=0}^{T-1}|\langle L,\Xi_n(0)\varphi_0\rangle|^2
	&=\left(\frac{1-\cos\omega}{3-2\cos\omega}\right)^2 \left( 1+\frac{\sin\omega}{1+\sin^2\omega} \right), \\
\overline{\mu}_\infty^{(\Psi_0,R)}(0)\equiv \lim_{n\to \infty}\frac{1}{T}\sum_{n=0}^{T-1}|\langle R,\Xi_n(0)\varphi_0\rangle|^2
	&=\left(\frac{1-\cos\omega}{3-2\cos\omega}\right)^2 \left( 1-\frac{\sin\omega}{1+\sin^2\omega} \right). 
\end{align*}
Note that $\overline{\mu}_\infty^{(\Psi_0)}(0)=\overline{\mu}_\infty^{(\Psi_0,L)}(0)+\overline{\mu}_\infty^{(\Psi_0,R)}(0)$. 
Finally we show that 
a stationary measure with some appropriate parameters is identical with the time averaged limit measure in Theorem \ref{acco1}. 
\begin{pro}
We consider a QW with one defect whose evolution is determined by $U^{(s)}$. 
Suppose that the two parameters in Proposition \ref{dis} satisfy 
\begin{align*} 
|\phi_0^{(L)}|^2=\overline{\mu}_\infty^{(\Psi_0,L)}(0),\;\;|\phi_0^{(R)}|^2=\overline{\mu}_\infty^{(\Psi_0,R)}(0), 
\end{align*}
respectively. Then we obtain 
\begin{equation}\label{hinano2}
\mu_I^{(\widetilde{\Psi}_0^{(\sigma,\tau)})}=\overline{\mu}_\infty^{(\Psi_0)}. 
\end{equation}
\end{pro}
We should remark that the evolution of the QW in LHS of Eq. (\ref{hinano2}) is the same as that in RHS.
The initial state in LHS corresponds to an eigenvector of $U^{(s)}$, which has an infinite support. 
On the other hand, the support of the initial state in RHS is just the origin. \\
From Eqs. (\ref{john10}) and (\ref{john11}), we easily find a general form of the time averaged limit measure of the walk, i.e., an exponential function 
skipping the computations imposed by Section 3. 
However when we require the values of two parameters $\phi^{(L)}(0)$ and $\phi^{(R)}(0)$ 
to obtain an explicit expression for the time averaged limit measure, 
after all we should come back again to the arguments of Section 3. 


\par
\
\par\noindent
{\bf Acknowledgment.} The authors thank K. Chisaki, T. Machida for helpful discussions and comments. 
NK acknowledges financial support of the Grant-in-Aid for Scientific Research (C) of Japan Society for the Promotion of Science (Grant No. 21540118). 
TL is partially supported by the Foundation for Polish Science. 
\par
\
\par
\bibliographystyle{jplain}

\begin{thebibliography}{99}
\bibitem{KonnoLocal}
Konno, N.: 
Localization of an inhomogeneous discrete-time quantum walk on the line. 
Quantum Inf. Proc.  \textbf{9} 405 (2010). 

\bibitem{KonnoRE} 
Konno, N.: 
One-dimensional discrete-time quantum walks on random environments. 
Quantum Inf. Proc. {\bf 8}, 387--399 (2009)

\bibitem{TregennaEtAl}
Tregenna, B., Flanagan, W., Maile, R. and Kendon, V.: 
Controlling discrete quantum walks: coins and initial states. 
New J. Phys. \textbf{5} (2003) 83. 

\bibitem{IKK}
Inui, N., Konishi, Y. and Konno, N.: 
Localization of two-dimensional quantum walks. 
Phys. Rev. A \textbf{69} 052323 (2004). 

\bibitem{ShikanoKatsura}
Shikano, Y. and Katsura, H.: 
Localization and fractality in inhomogeneous quantum walks with self-duality. 
Phys. Rev. E 82, 031122 (2010). 

\bibitem{JoyeP}
Joye, A. and Merkli, M.: 
Dynamical localization of quantum walks in random environments. 
J. Stat. Phys. \textbf{140} 1023-1053 (2010). 

\bibitem{AhlbrechtP}
Ahlbrecht, A., Scholz, V. B. and Werner, A. H.: 
Disordered quantum walks in one lattice dimension, 
J. Math. Phys. \textbf{52} 102201 (2011). 

\bibitem{W}
Wojcik, A., {\L}uczak, T., Kurzynski, P., Grudka, A. and Bednarska, M.: 
Quasiperiodic dynamics of a quantum walk on the line. 
Phys. Rev. Lett. \textbf{93} 180601 (2004).  

\bibitem{CGMV2}
Cantero, M. J., Gr\"unbaum, F. A., Moral, L. and Vel\'azquez, L.:
One-dimensional quantum walks with one defect. 
arXiv:1010.5762 (2010). 

\bibitem{BS}
Simon, B.:  
Orthogonal Polynomials on the Unit Circle Parts 1 and 2, 
American Mathematical Society Colloquim Publications, \textbf{54} (2005). 

\bibitem{CGMV}
Cantero, M. J., Gr\"unbaum, F. A., Moral, L. and Vel\'azquez, L.: 
Matrix-valued Szeg\H{o} polynomials and quantum random walks. 
Comm. Pure Appl. Math. {\bf 63}, 464--507 (2010)

\bibitem{KS2010}
Konno, N. and Segawa, E.: 
Localization of discrete-time quantum walks on a half line via the CGMV method. 
Quant. Inf. Comput. {\bf 11} 0485 (2011). 

\bibitem{LindenSharam}
Linden, N., Sharam, J.: 
Inhomogeneous quantum walks. 
Phys. Rev. A \textbf{80} 052327 (2009). 

\bibitem{Konno2002}
Konno. N.: Quantum random walks in one dimension. 
Quantum Inf. Proc. {\bf 1}, 345 (2002). 

\bibitem{Konno2005}
Konno, N.: A new type of limit theorems for the one-dimensional quantum random walk. 
J. Math. Soc. Jpn. 57, 1179.1195 (2005)

\bibitem{KonnoDis}
Konno, N.: 
A path integral approach for disordered quantum walks in one dimension. 
Fluctuation and Noise Letters, Vol.5, No.4, pp.529-537 (2005) 

\bibitem{RibeiroEtAl}
Ribeiro, P., Milman, P. and Mosseri, R.: 
Aperiodic quantum random walks. 
Phys. Rev. Lett. \textbf{93} 190503 (2004). 

\bibitem{MackayEtAl}
Mackay, T. D., Bartlett, S. D., Stephanson, L. T. and Sanders,  B. C.: 
Quantum walks in higher dimensions. 
Journal of Physics A: Math. Gen. \textbf{35} (2002) 2745. 

\bibitem{JoyeT}
Joye, A.: 
Random time-dependent quantum walks.  
arXiv:1010.4006 (2010).  

\bibitem{AhlbrechtT}
Ahlbrecht, A., Vogts, H., Werner, A. H. and Werner, R. F.: 
Disordered quantum walks in one lattice dimension. 
arXiv:1009.2019 (2010). 

\bibitem{Chandrashekar}
Chandrashekar, C. M.:
Disordered quantum walk-induced localization of a Bose-Einstein condensate. 
Phys. Rev. A \textbf{83} (2010) 022320. 

\bibitem{Machida}
Machida, T.: 
Limit theorems for a localization model of 2-state quantum walks, 
Int. J. Quantum Inf. \textbf{9} (2011) 863. 

\bibitem{IK}
Inui, N. and Konno, N.:  
Localization of Multi-State Quantum Walk in One Dimension. 
Physica A \textbf{353} (2005) 133.

\bibitem{IKS}
Inui, N., Konno, N. and Segawa, E.: 
One-dimensional three-state quantum walk, 
Phys. Rev. E \textbf{72} (2005) 056112.

\bibitem{FKK}
Katori, M., Fujino, S. and Konno, N.: 
Quantum walks and orbital states of a Weyl particle. 
Phys. Rev. A \textbf{72} (2005) 012316. 

\bibitem{MKK}
Miyazaki, T., Katori, M. and Konno, N.:
Wigner formula of rotation matrices and quantum walks. 
Phys. Rev. A \textbf{76} (2007) 012332.

\bibitem{SK}
Segawa, E. and Konno, N.: 
Limit theorems for quantum walks driven by many coins. 
Int. J. Quantum Inf. \textbf{6} 1231 (2008). 

\bibitem{KM}
Konno, N. and Machida, T.: 
Limit theorems for quantum walks with memory. 
Quant. Inf. Comput. \textbf{10} 1004 (2010). 

\bibitem{LP}
Liu, C. and Petulante, N.: 
One-dimensional quantum random walk with two entangled coins. 
Phy. Rev. A \textbf{79} 032312 (2009). 

\bibitem{ChaobinPetulante}
Liu, C. and Petulante, N.: 
On limiting distributions of quantum Markov chains. 
arXiv:1010.0741 (2010). 

\bibitem{KMcG}
Karlin, S., McGregor, J.: Random walks, 
Illinois J. Math. \textbf{3} 66--81 (1959). 

\bibitem{CHKS} 
Chisaki, K., Hamada, M., Konno, N. and Segawa, E.: 
Limit theorems for discrete-time quantum walks on trees. 
Interdiscip. Inform. Sci. {\bf 15} 423--429 (2009)

\bibitem{CKS}
Chisaki, K., Konno. N. and Segawa, E.: 
Limit theorems for the discrete-time quantum walk on a graph with joined half lines. 
arXiv:1009.1306 (2010). 

\bibitem{Flajolet}
Flajolet, P. and Sedgewick, R.: 
Analytic Combinatorics. Cambridge University Press (2009). 



\end{thebibliography}


\end{document}